\documentclass[epj]{svjour}
\usepackage{graphicx}
\usepackage{bm}   
\usepackage{color}
\usepackage{amssymb}
\usepackage{amsmath}
\usepackage{balance}
\usepackage{subfigure}

\begin{document}
	
\title{Extreme bursting events via pulse-shaped explosion in mixed Rayleigh-Li\'enard nonlinear oscillator}
\author{B. Kaviya \and R. Suresh \and V. K. Chandrasekar \mail{suresh@eee.sastra.edu}
}                     
%
%
\institute{Department of Physics, Centre for Nonlinear Science and Engineering, School of Electrical and Electronics Engineering,\\ SASTRA Deemed University, Thanjavur 613 401, India.}
\date{Received: -- / Revised version: --}
%
\abstract{
	We study the dynamics of a parametrically and externally driven Rayleigh-Li\'enard hybrid model and report the emergence of extreme bursting events due to a novel pulse-shaped explosion mechanism. The system exhibit complex periodic and chaotic bursting patterns amid small oscillations as a function of excitation frequencies. In particular, the advent of rare and recurrent chaotic bursts that emerged for certain parameter regions is characterized as extreme events. We have identified that the appearance of a sharp pulse-like transition that occurred in the equilibrium points of the system is the underlying mechanism for the development of bursting events. Further, the controlling aspect of extreme events is attempted by incorporating a linear damping term, and we show that for sufficiently strong damping strength, the extreme events are eliminated from the system, and only periodic bursting is feasible.
	\PACS{
		{PACS-key}{Rayleigh-Li\'enard model \and Extreme bursting \and Pulse-shaped explosion \and Chaotic bursting patterns \and Extreme events}
	} 
}
\titlerunning{Extreme bursting events via pulse-shaped explosion in mixed Rayleigh-Li\'enard nonlinear oscillator}
\authorrunning{B. Kaviya et al.}
\maketitle
\section{Introduction}
\label{sec:1}
The abrupt emergence of fast, repetitive large-amplitude peaks (spiking or active state) alternate with the slow, steady-state like small-amplitude oscillations (quiescent or rest state) at regular or irregular time intervals are characterized as bursting. Such complex and multiple-time scale oscillatory behavior is often appeared in biological systems, especially in neuronal systems \cite{terman1992,rowat2004,laing2003}, but also reported in physical, chemical, mechanical, and other systems \cite{simo2011,fetzer2018,nahmias2020,nowacki2012,brons1991}. Due to the universal occurrence of this phenomenon, the concept of bursting and its emerging dynamical mechanisms has been intensively studied in mathematical models \cite{deng1994,plant1981,deveris2001,medvedev2006,channell2007,hens2015,ghosh2009}, and in various real-world systems \cite{organ2003,meucci2002,makouo2017,yang2021,li2021}. Accordingly, based on the multiple-scale analysis, especially, using fast-slow subsystem analysis  \cite{kuehn2015,han2015,bertram2017}, it has been demonstrated that bursting is often emerged in the dynamical systems due to various mechanisms such as singular Hopf bifurcation  \cite{curtu2010}, canard phenomenon  \cite{desroches2013,krupa2001,han2012}, the breakup of invariant torus  \cite{larter1991,han2010}, via boundary crisis  \cite{han2017a,han2017b}, the blue-sky catastrophe  \cite{shilnikov2005}, through delayed bifurcation   \cite{han2016,yu2016,wang2020}, speed escape of attractors  \cite{han2017c}, via pulse-shaped explosion  \cite{han2018,wei2021,ma2021}, and few other dynamical mechanisms   \cite{vo2016,ryashko2017,han2017d,chen2017,buchele2006}. In particular, very recently, an interesting sharp transition behavior called pulse-shaped explosion has been reported in Rayleigh's type nonlinear dynamical system, in which the periodic complex bursting patterns occurred due to the appearance of pulse-shaped sharp quantitative changes in the branches of the equilibrium point and limit cycle attractor  \cite{han2018,wei2021}. In continuation with this, bursting oscillations induced by the bistable pulse-shaped explosion are also demonstrated in a nonlinear oscillatory system using multiple slow excitation frequencies  \cite{ma2021}. Two different types of regular bursting patterns, that is, bursting types of point-point type and cycle-cycle type, have been observed in the Rayleigh system via pulse-shaped explosion  \cite{han2018}. In addition to that, the pulse-shaped explosion can also induce different bursting patterns in other dynamical systems as well  \cite{song2020,chen2020}. Even though considerable work has been done in recent times regarding the pulse-shaped explosion-induced bursting dynamics, this area still requires further investigation. 

Recently, the emergence of extreme events (EE), defined as rare, recurrent, and large amplitude oscillations (events) that are significantly larger than the normal small amplitude oscillations, are manifested in state variables or observables of dynamical and real-world systems. This typical phenomenon has appeared in many natural and artificial systems without prior warning, but it greatly impacted life and society. Considering the widespread impact, EE have been documented in various form of disasters in natural systems like floods  \cite{buchele2006}, volcanic eruptions  \cite{sachs2012}, tsunamis  \cite{bird2011}, droughts  \cite{hoerling2013}, regime shifts in ecosystems  \cite{scheffer2003}, epidemic spreading  \cite{mcmichael2015,machado2020}, epileptic seizure in the human brain   \cite{lehnertz2006,frolov2019}, solar flares  \cite{buzulukova2017}, and so on. Apart from that EE also appeared in the form of share market crashes  \cite{krause2015}, mass panics  \cite{helbing2001}, jamming in computer and transportation networks  \cite{zhao2005,chen2015}, large power black outs  \cite{dobson2007}, collapse of large buildings due to extreme loading conditions  \cite{almarwae2017,suresh2021}, and many more. These examples motivate researchers from different scientific domains to investigate such phenomena from statistical and dynamical system aspects. 

The trademark characteristic properties of EE are unpredictability, and irregular occurrence, which is also the essential features of chaos. Thus, researchers are curious to observe the evidence of EE in nonlinear dynamical systems. Based on this, from the EE perspective, several motivations have emerged in the dynamical systems. Especially, understanding the development of extreme events in dynamical systems, their emerging mechanism, and controlling aspects are rigorously studied (both in isolated and coupled systems) in recent literature (For additional reading, see ref   \cite{chowdhury2021} and the references therein). {\bf In continuation with that, several novel routes to EE emergence have been demonstrated in dynamical systems. For example, the emergence of optical rogue waves in an optically injected laser system is experimentally demonstrated  \cite{zamora2013}, in which EE emerged in the system via an external crisis-like route. Similarly, EE can also appear in the system via interior and boundary crisis as well  \cite{ray2019,bonatto2017,kaviya2020}. Several studies reported that EE has also emerged in the system through the intermittency route   \cite{suresh2020,suresh2018,kingston2017}. Another route to the emergence of EE in systems with discontinuous boundaries is through sliding bifurcation   \cite{kumarasamy2018}, which has been reported recently. Extreme events originate in the systems due to instability in antiphase synchronization of the coupled systems via two different routes, intermittency and quasiperiodicity routes to complex dynamics for purely excitatory and inhibitory chemical synaptic coupling, respectively  \cite{mishra2018}. The authors in ref.  \cite{chowdhury2019} presented the origination of EE in coupled moving agents model due to the instability of the synchronization manifold. So far, we have discussed the emergence of EE in deterministic systems, which have a self-sustaining mechanism for generating EE under various circumstances. However, an important class of rare EE is induced by noise  \cite{pisarchik2011,pisarchik2012}. But, the route the system takes during each transition to EE in stochastic systems is not as clear as in deterministic systems. 
	
Despite a large number of recent publications reporting the emerging routes of EE, the subject is still young and contains interesting problems like the possibility of exploring new mechanisms and routes to EE in nonlinear dynamical systems is still open that demand the attention of researchers. In particular, the recently reported pulse-shaped explosion route to complex bursting is based on a specific Rayleigh oscillator. But a significant number of hybrid systems related to nonlinear oscillators of Rayleigh's type are available in literature  \cite{kanai2012,erlicher2013}, which are used to describe many physical phenomena and processes in electrical and mechanical systems  \cite{warminski2010,szabelski1995,hasegawa2011,miwadinou2018}. In addition to this, it is also of interest to investigate how systems with aperiodic dynamics, especially chaotic systems, respond to this pulse-shaped explosion is worth exploring.
	
Taking this as our motivation, in the present paper, we study the dynamics of a hybrid Rayleigh-Li\'enard model driven by external and parametric periodic excitations and explore the emerging dynamics as a function of the excitation frequencies. We observe that the system shows the possibility of the development of periodic and chaotic bursting that appeared due to a sharp explosion in the equilibrium points of the system, known as a pulse-shaped explosion. Although the emergence of periodic compound bursting patterns via pulse-shaped explosion is reported, earlier  \cite{han2018,song2020,chen2020}, the generic nature of the pulse-shaped explosion induced bursting pattern is demonstrated in the chosen model. Also, interestingly, we provide evidence for the development of rare but recurrent chaotic large bursting along with regular small oscillations, which approved all the conditions of the EE phenomenon. Hence, we discovered a novel route to EE in the nonlinear system, and it is worth mentioning that, to the best of our knowledge, this is the first study to report the emergence of EE via a pulse-shaped explosion in a dynamical system. The pulse-shaped explosion is confirmed using the transformed fast-slow system of the hybrid Rayleigh-Li\'enard model with a single slow variable adjusted to characterize the bursting oscillations. Further, in the extreme event literature, the threshold height is primarily used to distinguish EE from other small oscillatory events. However, we pointed out that characterizing EE based on the threshold height is not adequate for the systems, particularly those exhibits periodic and chaotic bursting. Therefore, additional statistical techniques are required to differentiate the rare events from the repetitive large bursting patterns. Consequently, the rarity of the chaotic bursting is verified using the histogram of inter-event intervals, which satisfies the Poisson-like distribution. Further, we investigate the influence of linear damping on EE. We found that the chaotic oscillations tamed into periodic bursting while increasing the damping strength. Consequently, the region of EE in parameter space is slowly reduced as a function of damping strength. Thus, the hybrid system proposes mitigating chaotic bursting events in the form of EE using linear damping.}

The remainder of the paper is organized as follows: In Sec.\ref{sec:2}, the model we have taken for the study is introduced. The emergence of complex periodic and chaotic bursting patterns in the hybrid model is demonstrated in Sec.\ref{sec:3}. Also, the fast-slow system theory is introduced to confirm the pulse-shaped explosion route. Further, in Sec.\ref{sec:4}, the controlling aspect of EE is investigated by presenting a linear damping term in the model equation. Finally, the results are summarized and concluded in Sec.\ref{sec:5}.
\begin{figure}[!ht]
\begin{center}
\includegraphics[width=0.7\textwidth]{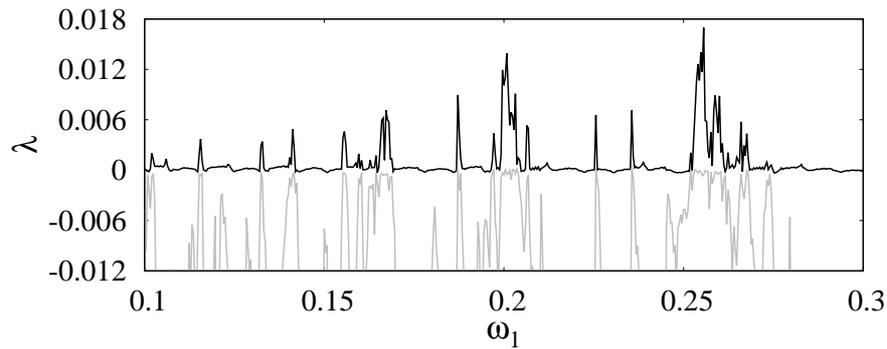}
\end{center}
\caption{The Lyapunov exponents of the system (\ref{eqn1}) is plotted as a function of $\omega_1$ shows the chaotic dynamics of the system for various values of $\omega_1$. The other system parameters are fixed at $\alpha=-0.25$, $\beta= 0.05$, $\gamma= 0.1$, $\omega_2=0.04$, $F$=0.08, and $\mu= 0.99$.}
\label{fig1}
\end{figure} 

\section{Model equation}
\label{sec:2}
To demonstrate the results, we consider a mathematical model of a mixed Rayleigh-Li\'enard oscillator with external and parametric periodic excitations. The equation of motion of the hybrid model is given by  
\begin{equation}
\ddot{x}+f(x,\dot{x})+ [1-\mu \cos(\omega_1 t)] (x+\gamma x^3) = F \cos(\omega_2 t),
\label{eqn1}
\end{equation}

here, the function $f(x,\dot{x}) = -\alpha x \dot{x}+\beta \dot{x}^{3}$, in which $\alpha$, and $\beta$ indicates the quadratic and cubic nonlinear damping coefficients of Rayleigh-Li\'enard type, $\gamma$ is the nonlinear stiffness constant, $\mu$ and $F$ are the amplitudes of the external and parametric excitations, and $\omega_1$ and $\omega_2$ are the frequencies of the external and parametric excitations, respectively. The class of mixed Rayleigh-Li\'enard system is widespread in the areas of science, and engineering. The dynamics of this hybrid system have been intensively studied in the literature, and exciting results such as period-doubling bifurcation leading to chaotic dynamics, strange attractors, symmetry breaking, and so on have been reported using various excitation models  \cite{miwadinou2018,margallo1992,maccari2001,kpomahou2021}. Also, the quadratic nonlinear damping term $\alpha x \dot{x}$ appeared in many real-world models and was found to play a crucial role mostly in microelectromechanical, and nanoelectromechanical oscillators   \cite{eichler2011,zaitsev2012}, fluid mechanics  \cite{ran2009}, and have implications in mass and force sensing applications  \cite{ekinci2004,papariello2016}, mechanical noise squeezing in laser cooling technology   \cite{akerman2010}.

\begin{figure}[!ht]
\centering
\includegraphics[width=1.0\textwidth]{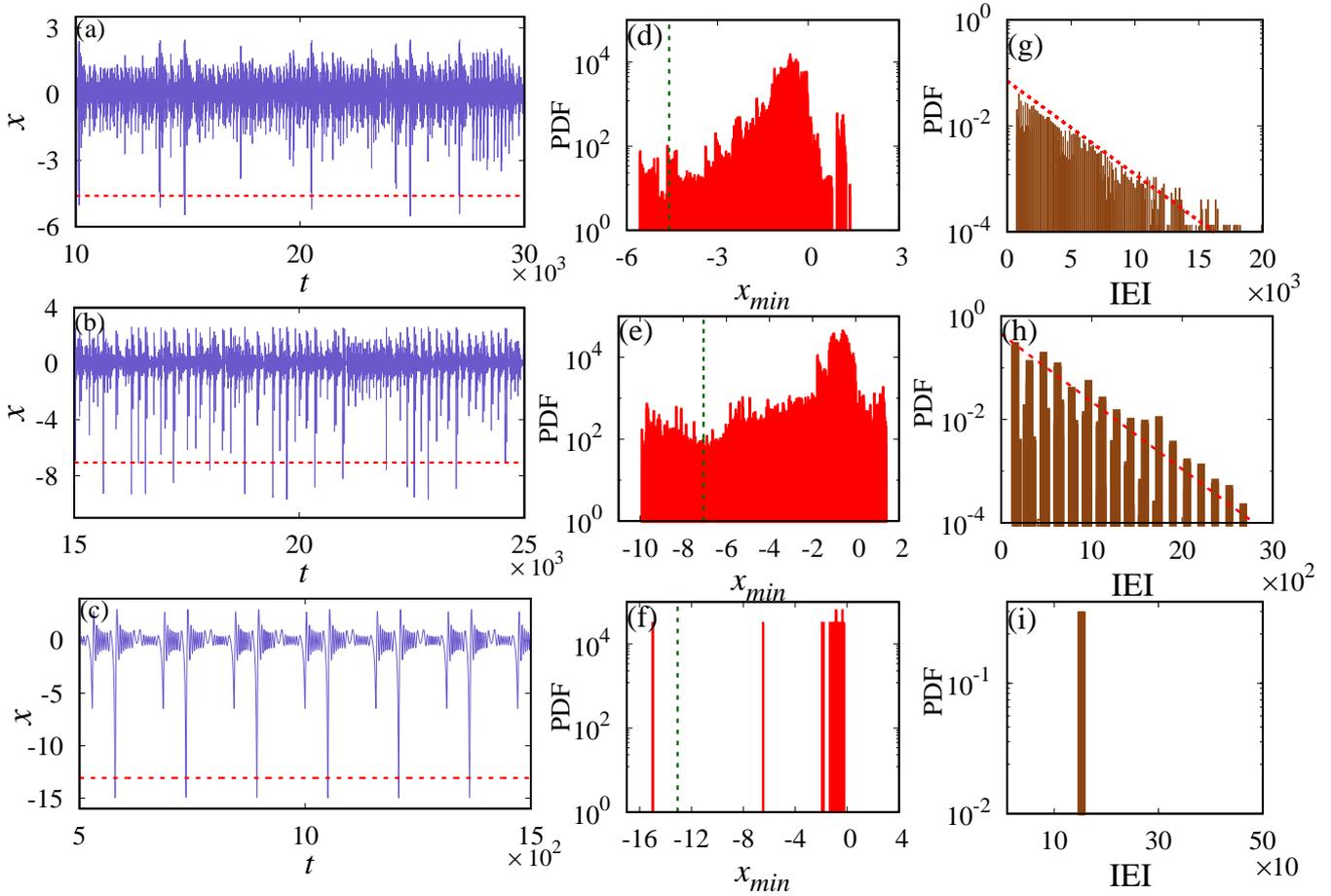}
\caption{Left panel from top to bottom displays the time evolution of the state variable ($x$) of the system (\ref{eqn1}) exhibits the emergence of chaotic and periodic bursting oscillations for different $\omega_1$ values. The dashed horizontal line represents the threshold height ($H_T$). The middle panel shows the probability distribution function of the respective time series data and threshold height, $H_T$ (vertical dashed line). The right panel portrays the probability distribution function of the inter-event interval, and the dashed line represents the Poisson distribution. (a, d, g), and (b, e, h) displays the emergence of chaotic bursting patterns with infrequent large oscillations characterized as EE for $\omega_1=0.2$ and $0.166$, respectively. (c, f, i) shows the emergence of regular periodic bursting patterns for $\omega_1=0.12$. Other system parameters are fixed as given in Fig. \ref{fig1}.}
\label{fig2}
\end{figure}
\section{Emergence of complex bursting patterns}
\label{sec:3}
The hybrid Rayleigh-Li\'enard system (\ref{eqn1}) exhibits periodic and chaotic behavior as a function of the frequencies of external and parametric excitations for fixed values of system parameters. In order to study the dynamics of the system, Eq. (\ref{eqn1}) is numerically integrated using the fourth-order Runge-Kutta method with the time step of 0.01. For numerical integration, the system parameters are fixed at $\alpha=-0.25$, $\beta=0.05$, $\gamma=0.1$, $\mu=0.99$, $\omega_2=0.04$, and $F=0.08$ throughout the manuscript until and otherwise mentioned. 

The emergence of periodic and chaotic states of the system is quantified using the Lyapunov exponents ($\lambda$) by varying the parametric excitation frequency in the range of $\omega_1\in(0.1, 0.3)$ by fixing the other parameters as given above. The estimated Lyapunov exponent is depicted in Fig. \ref{fig1}. The positive peaks ($\lambda>0$) of the first Lyapunov exponent (black line) indicate the emergence of chaos, and $\lambda<0$ (light gray line) shows the periodic dynamics. Notably, both chaos and periodic oscillations appeared in the form of bursting for the selected parameter region. The qualitative validation of those bursting can be seen from the time snaps of the system for different $\omega_1$ values. 

The time evolution of the system (\ref{eqn1}) is depicted in the left panel of Fig.~\ref{fig2} for three different values of $\omega_1$ (by decreasing $\omega_1$ from higher to lower values) to demonstrate the bursting patterns. To be precise, for $\omega_1=0.2$, the system manifests chaos, in which intermittent, rare, and large-amplitude peaks have appeared at random intervals in the negative $x$ values amid small-amplitude oscillations, which are characterized as extreme chaotic bursts or named as EE. The reason for bursting in the negative $x$ values will be discussed later. The time snaps consist of rare and recurrent large-amplitude oscillations along with the small-amplitude oscillations are portrayed in Fig. \ref{fig2}(a). The large excursions are clearly visible in the figure. Usually, those large excursions can be qualitatively distinguished from other peaks using the threshold height  \cite{chowdhury2021,suresh2018} $H_T = \langle P_{min}\rangle- N\sigma$, in which $\langle P_{min}\rangle$ is the time average of the local minima of the time series, $\sigma$ is the standard deviation and $N$ is an arbitrary integer, which varies based on the dynamical system. For system (\ref{eqn1}), we choose the value of $N=6$. The negative sign in the threshold height equation is due to the negative peaks that occurred in the system. In order to estimate the threshold height, we have used $2\times10^9$ data points after leaving adequate transient. The calculated value is marked as a horizontal dashed line in Fig. \ref{fig1}(a). The negative peaks lower than $H_T$ are characterized as EE. A similar type of extreme dynamical state can also be observed in the system for several other values of $\omega_1$ when the system exhibits chaos. Figure \ref{fig2}(b) shows an another example for the emergence of EE for $\omega_1=0.166$. Compare to Fig.\ref{fig2}(a) the size of the attractor is enlarged in Fig. \ref{fig2}(b). Hence larger peaks are observed in the system. 

In contrast to the previous two cases, when we choose $\omega_1$ in the periodic regime, the system shows \emph{repetitive patterns} of small amplitude oscillations, which are alternated by large-amplitude bursting (peaks) at regular intervals of time characterized as periodic bursting. Time snaps of one such periodic state are observed for $\omega_1=0.12$ and portrayed in Fig. \ref{fig2}(c). When we estimate $H_T$ and plot it along with the time series, the prominent repetitive negative peaks are below the threshold height for the same value of $N$. However, they fail to prove EE's randomness and unpredictability since they are repetitive patterns. Thus, the verification of EE using the estimation of threshold height $H_T$ is not sufficient for systems that exhibit periodic and chaotic bursting patterns in different parameter regions.

Therefore, it is necessary to characterize and confirm EE using other statistical techniques. Hence, we have estimated the probability distribution function (PDF) of the corresponding time series data shown in Figs. \ref{fig2}(a), (b), and (c) to confirm and distinguish EE from periodic bursting states and plotted in Figs. \ref{fig2}(d), (e), and (f), respectively. The PDF of the time series data provided in Figs. \ref{fig2}(a) and (b) show the non-Gaussian, continuous long-tail distribution corroborating the occurrence of EE. The vertical dashed line in Figs. \ref{fig2}(d) and (e) indicates $H_T$. On the other hand, the PDF of the time series data respective to the periodic oscillations (Fig. \ref{fig2}(c)) did not show the continuous long-tail distribution. Still, it displayed a single peak of the repetitive bursting. 

Further, the rarity of the events is quantified by the distribution of inter-event intervals ($IEI_m = t_{m+1}-t_m$), $m=1,2,\cdots,(M-1)$ which follows Poisson-like distribution when the system exhibit EE. Here $t_m$ is the time of the occurrence of $m^{th}$ event in a set of $M$ events. Figures \ref{fig2}(g), and \ref{fig2}(h) depicts the semi-log scale histogram plots of the estimated IEI for the time series of Figs. \ref{fig2}(a), and \ref{fig2}(b), respectively, satisfying the Poisson-like distribution $P(x) = \Lambda e^{-\Lambda x}$ with the scaling parameter $\Lambda=0.0004248$, and $\Lambda=0.0003625$ (dashed line), respectively. Nevertheless, the histogram of inter-event interval calculated for the periodic bursting failed to show the Poisson-like distribution due to the periodic nature of the system, which is evident from Fig.\ref{fig2}(i). Hence, we propose that estimating the threshold height alone is ineffective in recognizing EE in a system that exhibits periodic and chaotic bursting. Additionally, one inevitably evaluates other statistical techniques like calculating PDF and IEI helps to distinguish EE better from other normal events. 

\begin{figure}[!ht]
\begin{center}
\includegraphics[width=0.8\textwidth]{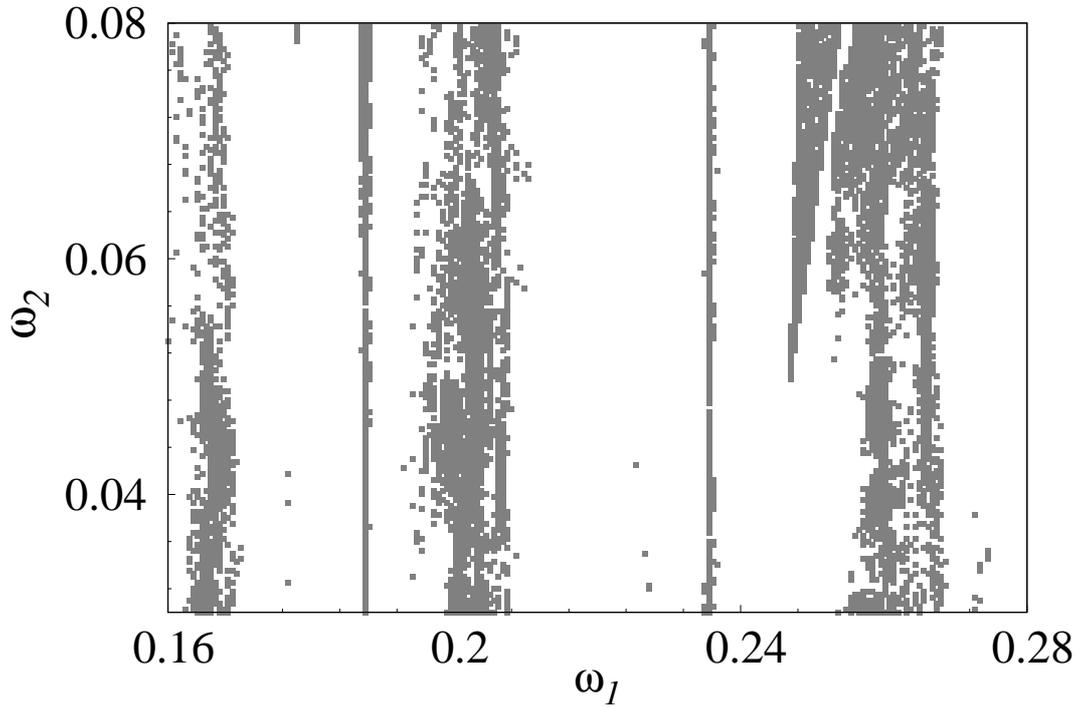}
\end{center}
\caption{The two parameter diagram in the parameter space of $\omega_1\in(0.16, 0.28)$ and $\omega_2\in(0.03, 0.08)$ shows the emergence of different bursting states. The dark gray points represent the parameter values to which the system exhibits EE, and white region indicates the periodic bursting states.}
\label{fig3}
\end{figure}

The global picture of the regions of periodic states and chaotic bursting in the form of EE is identified in the parameter space of two excitations frequencies $\omega_1\in(0.16, 0.28)$ and $\omega_2\in(0.03, 0.08)$, respectively, which is depicted in Fig. \ref{fig3}. As mentioned earlier, EE is exhibited when the system manifests chaos. Therefore, we assume (and confirm) that simultaneously estimating the Lyapunov exponent and the threshold height, $H_T$, helps us distinguish the EE region from the periodic bursting region as a function of $\omega_1$ and $\omega_2$. If the system shows chaos with a positive Lyapunov exponent, and when the minima of the $x$-variable are beneath $H_T$, then the dynamics that occurred at the specific frequencies are marked as EE (dark gray points in Fig. \ref{fig3}). Otherwise, the parameter values to which the Lyapunov exponent becomes negative are considered periodic bursting regions (the white region in Fig. \ref{fig3}). We also double-checked the emergence of those two bursting states in the specified parameter range.

It is interesting and crucial to understand the mechanism responsible for the emergence of such bursting events. To this purpose, we transform Eq. (\ref{eqn1}) into slow-fast system with one single slow variable $\delta(t) = \cos(\omega_2 t)$   \cite{han2018a}. For our study, the frequency ratio between the parametric and external excitations is fixed as $1:n$. That is, $\omega_1=n\omega_2$, in which $n$ is an odd integer. The modified equation for the fast subsystem of Rayleigh-Li\'enard system can be given as

\begin{equation}
\ddot{x}+f(x,\dot{x})+[1-\mu k_n(\delta)] (x+\gamma x^3) = A \delta,
\label{eqn2}
\end{equation}

in which $k_n(x)$ is the trigonometric polynomial for $\cos(n\omega_2 t)$ which can be  extended as $k_n(x) = D^0_nx^n - D^2_nx^{n-2} (1-x^2)+ D^4_nx^{n-4}(1-x^2)^2 - ...+(-1)^\frac{m}{2}D^m_nx^{n-m}(1-x^2)^\frac{m}{2}$, where $m$ is the maximum even number ($m<n$). In Eq. (\ref{eqn2}) when $1-\mu k_n(\delta)\neq0$, the fast subsystem has one stable equilibrium point ($x,0$) in which $x$ is the real root of $ [\mu k_n(\delta)-1](x+\gamma x^3)+A\delta=0$. When $1-\mu k_n(\delta)\rightarrow0$ the equilibrium point approach to infinity (critical transition) in accordance with the value of $\mu$ and $\delta$.

\begin{figure}[!ht]
\begin{center}
\includegraphics[width=0.6\textwidth]{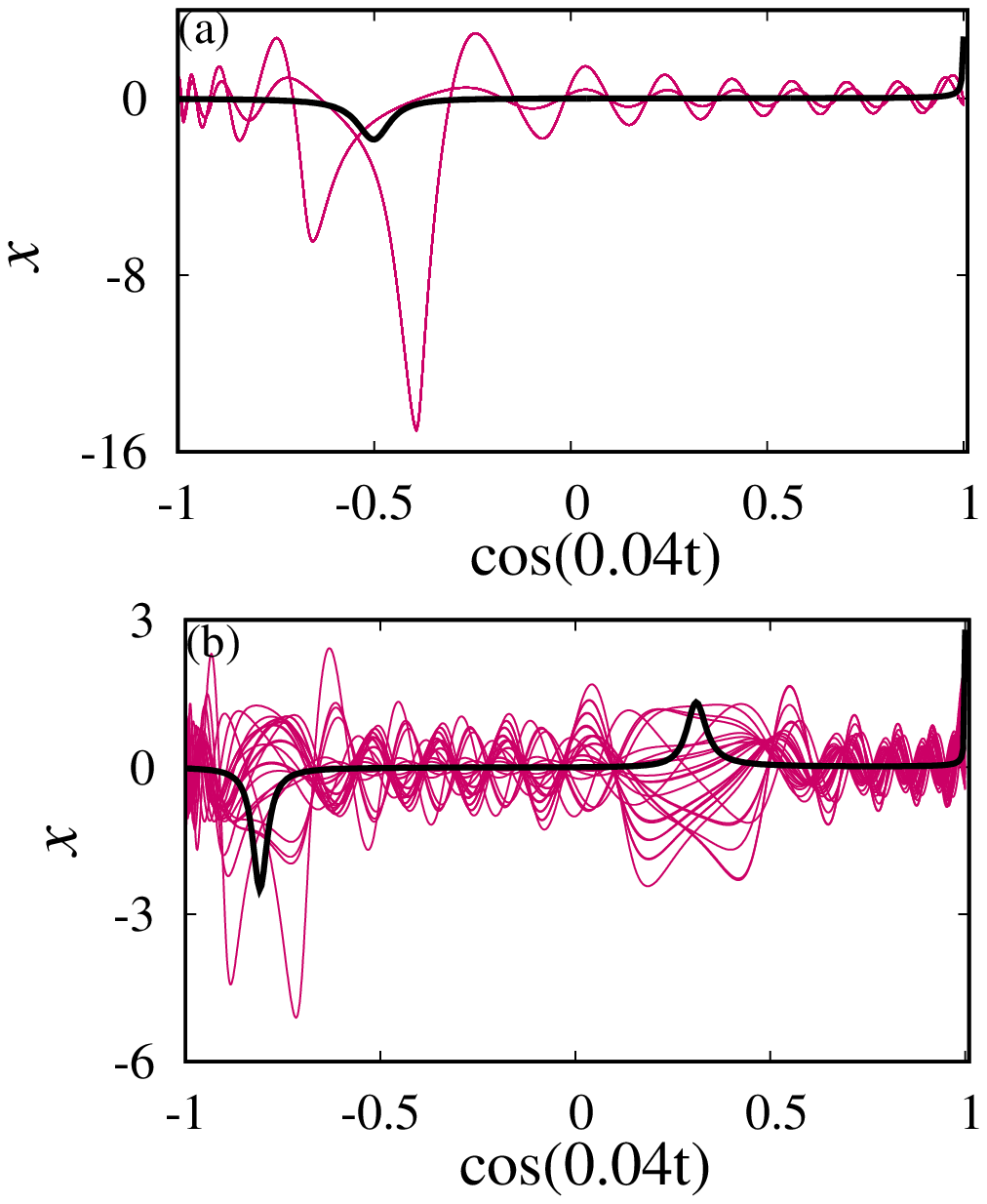}
\end{center}
\caption{Pulse shaped explosion related to (a) $n$ = 3, and (b) $n$ = 5 are plotted along with the bursting patterns occurred in the fast-slow subsystem (\ref{eqn2}) as a function of $\delta(t) = \cos(0.04t)$.}
\label{fig4}
\end{figure}

This critical transition drives the attractors to infinity in a narrow parameter space near the critical escape line at $\mu=1$. The birth and disappearance of equilibrium points in the form of a sharp pulse-shaped transition occurred in the system as a function of time, and several occurrences of this transition are based on the value of $n$, named as pulse-shaped explosion  \cite{han2018}. For an illustration, when $n=2$, Eq. (\ref{eqn2}) exhibits relatively trivial periodic bursting patterns. Nevertheless, taking $n=3$ as an example, we interpret the simple pulse-shaped explosion as showing a single peak. The black line in Fig. \ref{fig4}(a) shows the bifurcation of a stable equilibrium branch, which approaches infinity when $\mu$ is slightly less than 1 ($\mu=0.99$) at $\delta=\delta_c=-0.5$. Beyond this critical value of $\delta$, the two fragmented steep branches coalesce together and create a stable equilibrium branch again as a function of $\delta$. This process is repeated in the time domain. The establishment of such steep branches forms pulse-like sharp quantitative changes as a function of $\delta$, a pulse-shaped explosion. In other words, the steep region of the curve indicates the active area of bursting, and the flat area represents the rest area of bursting. Therefore, the system trajectory approaches near the steep region, making large excursions in the phase space and exhibiting large oscillations. The above mentioned case is shown in Fig. \ref{fig4}(a) for $n$ = 3. The red (gray) line depicts the trajectory of the fast-slow subsystem and shows the emergence of repetitive periodic bursting equivalent to Fig. \ref{fig2}(c). When we increase the $n$ value, the number of explosions is increased in the fast subsystem. We choose $n=5$ as an example of two peaks appearing in the pulse-shaped explosion for different $\delta$ values. The system exhibits chaos for the chosen parameter values and leads to large-amplitude irregular bursting when the trajectory goes near the steep region. Also, the amplitude of the large oscillations depends on how near the trajectory goes to the steep region, which occurs completely at random intervals of time due to the chaotic nature of the system. The small-amplitude oscillations appear in the flat region, which is also evident from Fig. \ref{fig4}(b). Thus, the emergence of chaos during the pulse-shaped explosion is the mechanism of the appearance of EE in the system for $n=5$.

\begin{figure}[!ht]
\begin{center}
\includegraphics[width=0.6\textwidth]{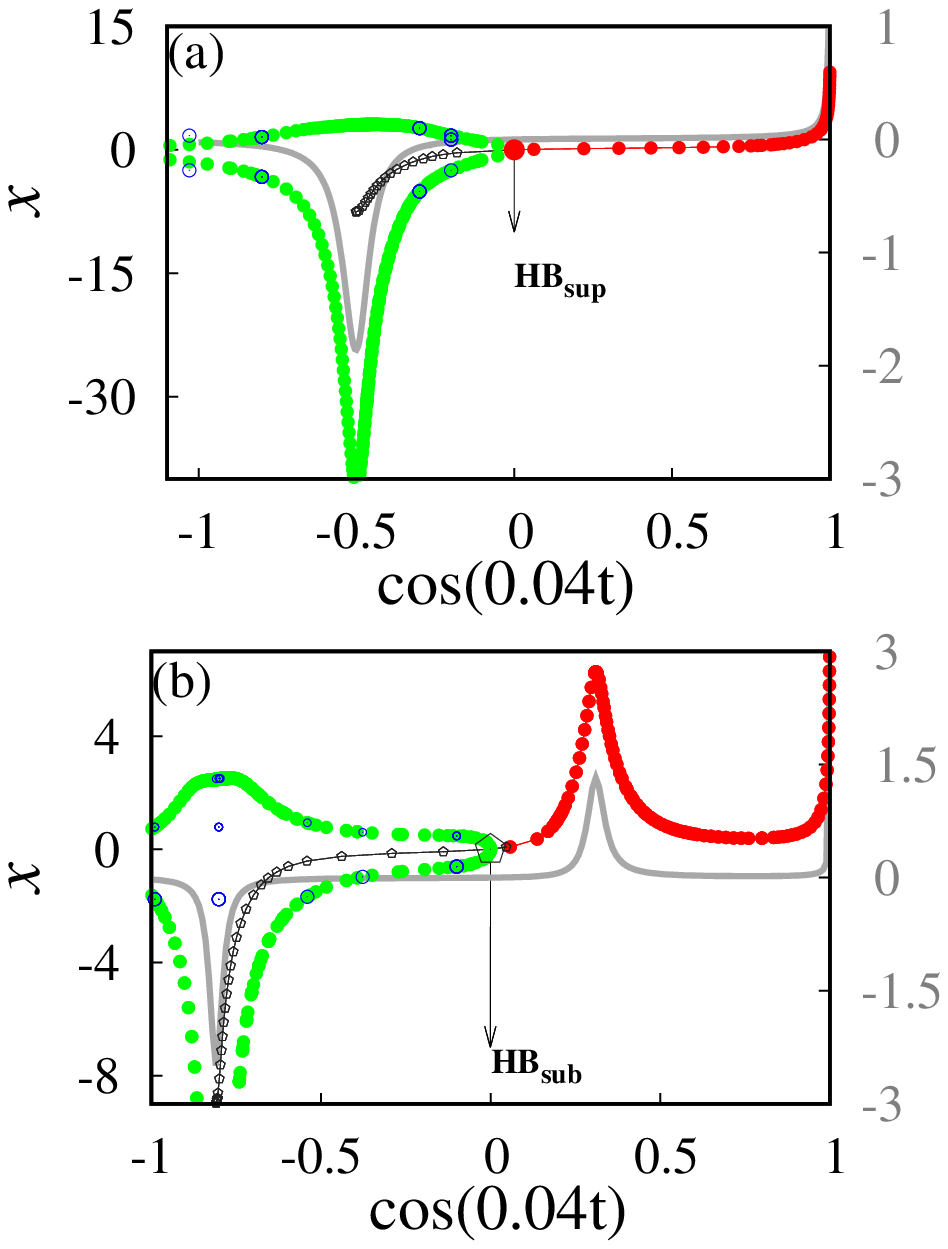}
\end{center}
\caption{Bifurcation of the equilibrium points resulting in complex bursting patterns as a function of $\cos(0.04t)$ for (a) $n$ = 3, and (b) $n$ = 5, respectively. Red (dark gray) and green (light gray) filled solid circles indicate stable equilibrium, and stable limit cycles, respectively. Black, and blue (dark gray) open circles denote unstable equilibrium and unstable limit cycle. {\bf The continuous gray line represents the pulse-shaped explosion curve and the right side values of the $y$-axis indicating the amplitude of the pulse-shaped explosion.}}
\label{fig5}
\end{figure}

One can notice that the periodic or chaotic bursting has occurred only in the negative region of $x$-variable even though the system exhibits pulse-shaped explosion both in positive and negative values of $x$-variable [Fig. \ref{fig4}]. In order to clarify this, we have determined the stability analysis of the equilibrium point using the XPPAUT as a function of $\delta(t)=\cos(0.04t)$ for two different cases of $n$=3, and 5, which are plotted in Fig. \ref{fig5}. In particular, Fig. \ref{fig5}(a) shows the bifurcation of equilibrium points for the case of $n=3$. When we look into the figure from right to left by decreasing the control parameter $\delta$ from 1 to -1, we observe the stable equilibrium point, marked as red (dark gray) filled circles, lost its stability at $\delta=0$ via a supercritical Hopf bifurcation. After that, the stable limit-cycle solution can be obtained for $\delta<0$, depicted by green (light gray) filled circles. One can note here that the stable limit-cycle always revolves around the unstable equilibrium point (marked as open black circles) and undergo a sharp transition in the negative $x$-direction at $\delta=-0.5$, which is the reason for the appearance of bursting in the negative $x$-direction.

Similarly, in the case of $n=5$, the stable equilibrium point produces a sharp transition in the positive $x$-direction for a specific value of $\delta>0$. Hence the chaotic trajectory crossing this $\delta$ region is slightly perturbed. However, it does not produce significant large oscillations in the positive $x$-direction. This is evident from Fig. \ref{fig4}(b). When we decrease $\delta$ further, the stable equilibrium is bifurcated into a stable limit-cycle orbit via a sub-critical Hopf bifurcation at $\delta=0$ (marked as $HB_{sub}$ in Fig. \ref{fig5}(b)). The emerged limit-cycle has a steep transition in the negative $x$-direction due to the pulse-shaped explosion, which leads the chaotic trajectory to exhibit large excursions in the negative direction as manifested in Fig. \ref{fig4}(b). The continuous gray line in Figs. \ref{fig5}(a) and \ref{fig5}(b) are pulse-shaped explosion plotted for comparison. 

Thus, we confirm that the system exhibit critical transition in the form of pulse shaped explosion, which is the mechanism for the emergence of bursting. Moreover, the emergence of chaotic oscillations is the reason for the emergence of EE in the Rayleigh-Li\'enard hybrid system. Further, the reason for the bursting in the negative $x$-direction is also explained. In the next section, we study the influence of linear damping on EE in the hybrid system.
\section{Influence of linear damping}
\label{sec:4}
\begin{figure}[!ht]
\includegraphics[width=1.0\textwidth]{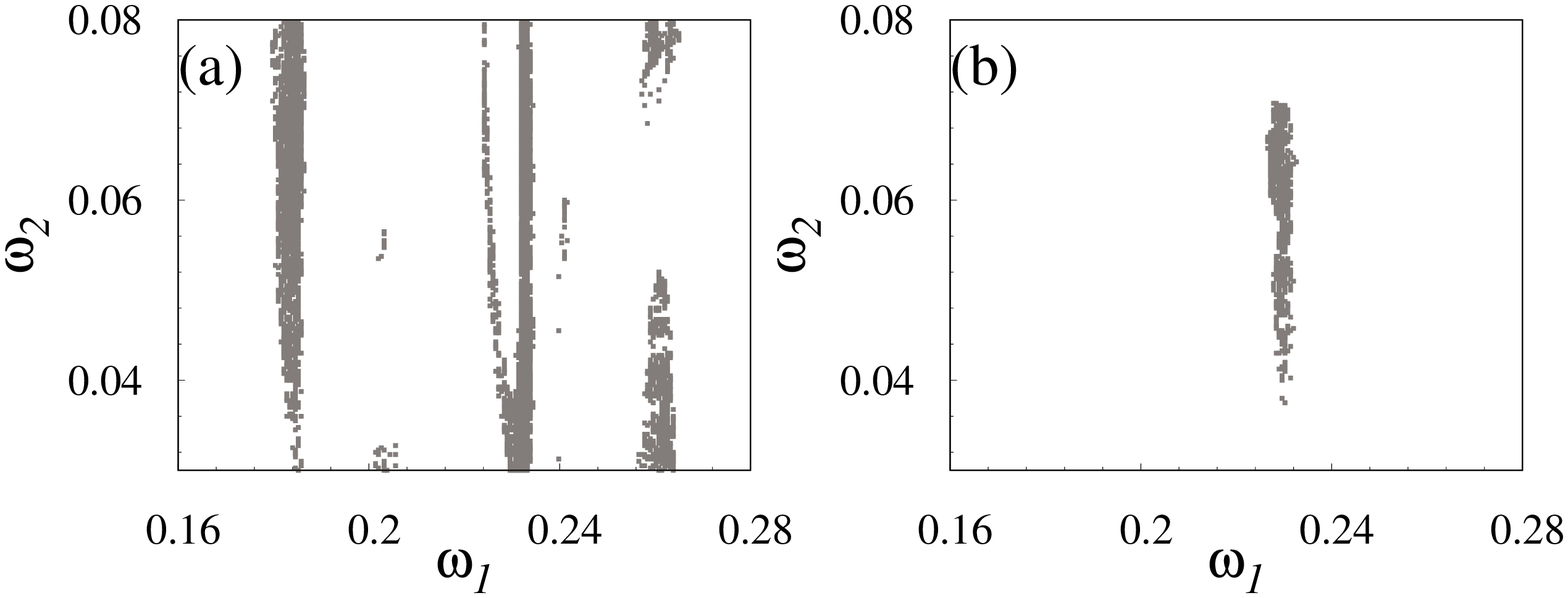}
\caption{The two parameter diagram of the system (\ref{eqn3}) plotted in the parameter space of $\omega_{1,2}$ parameter space for two different values of linear damping parameter (a) $\xi=0.03$, and $\xi=0.05$ depicts the attenuation of EE. The gray points shows the parameter values to which EE occurred and white region represents periodic bursting.}
\label{fig6}
\end{figure}

Controlling of EE is attempted by incorporating the linear damping term $\xi\dot{x}$ in Eq. (\ref{eqn1}) so that the equation can be rewritten as 

\begin{equation}
\ddot{x}+f(x,\dot{x})+\xi\dot{x}+ [1-\mu \cos(\omega_1 t)] (x+\gamma x^3) = F \cos(\omega_2 t),
\label{eqn3}
\end{equation}

here $\xi$ is the strength of linear damping. We observe that for a sufficient linear damping strength, the chaotic nature of the system is tamed, and the chaotic oscillations have vanished from the system. Thereby the chaotic bursting in the form of EE is eliminated from the system dynamics. In order to confirm the results we have estimated the emergence of EE as a function of $\omega_1$, and $\omega_2$ for two different values of $\xi$, which are depicted in Fig. \ref{fig6}(a) and \ref{fig6}(b) for $\xi$ = 0.03, and $\xi$ = 0.05, respectively. The gray points indicate the parameter values to which EE emerged in the system, and the white region indicates the periodic bursting oscillations. One can note that compared to Fig. \ref{fig3}, in Fig. \ref{fig6}(a), the gray shaded EE region is significantly reduced, confirming the elimination of EE. Further, increase in the linear damping to $\xi$ = 0.05, the region of EE is even more reduced, which is manifested in Fig. \ref{fig6}(b). Finally, when $\xi>$ 0.065, the system exhibits no EE, and only periodic oscillations are feasible. 
%

\section{Conclusion}
\label{sec:5}

We have systematically investigated the emerging bursting dynamics in the Rayleigh-Li\'enard hybrid system driven by parametric and external excitations to consolidate the results.

We found the emergence of periodic and chaotic bursting oscillations as a function of excitation frequencies. We have pointed out that bursting patterns occurred in the system due to the emergence of a sharp pulse-like explosion that appeared in the equilibrium points of the system, named the pulse-shaped explosion. Notably, during the emergence of chaos, large amplitude oscillations appeared in the system infrequently at random intervals. Those large-amplitude oscillations are characterized as EE. Even though the development of periodic bursting via the pulse-shaped explosion was reported earlier, this is the first study to document the development of EE via pulse-shaped explosion. Thus, a novel route to EE emergence is reported in a dynamical system. We also proposed that finding EE based on the threshold height is inadequate for the systems exhibiting periodic and chaotic oscillations. Therefore, additional statistical techniques are required to differentiate the rare events from the repetitive large amplitude oscillations. The rarity of the EE is verified using the inter-event interval histogram, which satisfies the Poisson-like distribution. 

We have identified that the simultaneous existence of chaos and the pulse-shaped explosion are critical mechanisms for the emergence of EE. The transformed fast-slow system with a single slow variable is used to characterize and validate the pulse-shaped explosion. Finally, the influence of linear damping on EE is studied. We found that the chaotic nature of the system is attenuated while increasing the damping strength, thereby eliminating EE from the system, and only periodic busting is feasible. 

This paper focuses on the emergence of extreme events via the pulse-shaped explosion in the Rayleigh-Li\'enard hybrid model related to the odd excitation frequency ratios. Exploring the pulse-shaped explosion route to EE in other dynamical systems with multiple-frequency slow excitations is also interesting. We are progressing our research in this direction. 

\color{black}
\section*{Acknowledgement}
B. Kaviya acknowledges SASTRA for providing Teaching Assistantship. The work of VKC is supported by the DST-CRG project under grant number CRG/2020/004353. All the authors thank DST, New Delhi, for computational facilities under the DST-FIST program with project number SR/FST/PS-1/2020/135 of the Department of Physics. \\

\noindent {\bf Author Contributions} All the authors contributed equally to the preparation of this manuscript.\\

\noindent {\bf Data Availability Statement} The authors confirm that the data supporting the findings of this study are available within the article.

\begin{thebibliography}{99}
	%
\bibitem{terman1992}
D.Terman, J. Nonlin. Sci \textbf{2}, (1992) 135–182.
%
\bibitem{rowat2004}
P.F. Rowat, R. C. Elson, J. Comput. Neuroci \textbf{16}, (2004) 87–112.
%
\bibitem{laing2003}
C.R. Laing, B. Doiron, A. Longtin, L. Noonan, R. W. Turner, L. Maler, J. Comput. Neuroci \textbf{14}, (2003) 329–42.
%
\bibitem{simo2011}
H. Simo, P. Woafo, Mech. Res. Commun \textbf{38}, (2011) 537–41.
%
\bibitem{fetzer2018}
F. Fetzer, J. Laser Appl \textbf{30}, (2018) 012009.
%
\bibitem{nahmias2020}
M. A. Nahmias, H-T. Peng, T. F. de Lima, C. Huang, A. N. Tait, B. J. Shastri, P.R. Prucnal, arXiv preprint arXiv:2012.08516 (2020).
%
\bibitem{nowacki2012}
J. Nowacki, H. M. Osinga, K. Tsaneva-Atanasova, J. Math. Neurosci \textbf{2}, (2012) 7.
%
\bibitem{brons1991}
M. Brons, K. Bar-Eli, J. Phys. Chem \textbf{25}, (1991) 8706-8713.
%
\bibitem{deng1994}
B. Deng B, Math. Biosci \textbf{119}, (1994) 241-250.
%
\bibitem{plant1981}
R. E. Plant, J. Math. Biol \textbf{11}, (1981) 15-32.
%
\bibitem{deveris2001}
G. de Veries, Phys. Rev. E \textbf{64}, (2001) 051914.
%
\bibitem{medvedev2006}
G.S. Medvedev, Phys. Rev. Lett \textbf{97}, (2006) 048102.
%
\bibitem{channell2007}
P. Channell, G. Cymbalyuk, A. Shilnikov A, Phys. Rev. Lett \textbf{98}, (2007) 134101.
%
\bibitem{hens2015}
C. Hens, P. Pal, S. K. Dana, Phys. Rev. E \textbf{92}, (2015) 022915.
%
\bibitem{ghosh2009}
A. Ghosh, D. Roy, V. K. Jirsa, Phys. Rev. E \textbf{80}, (2009) 041930.
%
\bibitem{organ2003}
L. Organ, I. Z. Kiss, J. L. Hudson, J. Phys. Chem \textbf{107},(2003) 6648-6659.
%
\bibitem{meucci2002}
R. Meucci, A.D. Garbo, E. Allaria E, F.T. Arecchi, Phys. Rev. Lett \textbf{88}, (2002) 144101.
%
\bibitem{makouo2017}
L. Makouo, P. Woafo, Chaos Soliton Fract \textbf{94}, (2017) 95-101.  
%
\bibitem{yang2021}
C. Yang, H. Hui, X. Song, S. Huang, J. Pressure Vessel Technol \textbf{143}, (2021) 031301.
\bibitem{li2021}
Z. Li, S.Fang, M. Ma, W. Wang, Int. J. Bifur. Chaos \textbf{31}, (2021) 2130023.
\bibitem{kuehn2015}
C. Kuehn, \textit{ Multiple Time Scale Dynamics} (Springer, Berlin 2015).
%
\bibitem{han2015}
X. Han, Q. Bi, P. Ji, J. Kurths, Phys. Rev. E \textbf{92}, (2015) 012911.
\bibitem{bertram2017}
R. Bertram, J. E. Rubin, Math. Biosci \textbf{287}, (2017) 105-121.
\bibitem{curtu2010}
R. Curtu, Physica D  \textbf{239}, (2010) 504-514.
%
\bibitem{desroches2013}
M. Desroches, T. J. Kaper, M. Krupa, Chaos \textbf{23}, (2013) 046106.
%
\bibitem{krupa2001}
M. Krupa, P. Szmolyan, Int. J. Differ. Equ \textbf{174}, (2001) 312-368.
%
\bibitem{han2012}
X. Han, Q. Bi, Nonlin. Dyn \textbf{68}, (2012) 275-283.
%
\bibitem{larter1991}
R. Larter, C.G. Steinmetz,Philos. Trans. Roy.
Soc. Lond. Ser. A \textbf{337},(1991) 291-298.
%
\bibitem{han2010}
X. Han, B. Jiang, Q. Bi, Nonlin. Dyn \textbf{61}, (2010) 667-676.
\bibitem{han2017a}
X. Han, C. Zhang C, Y. Yu, Q. Bi,  Int. J. Bifur. Chaos \textbf{27}, (2017) 1750051.
%
\bibitem{han2017b}
X. Han, Y. Yu, C. Zhang, Nonlin. Dyn \textbf{88}, (2017) 2889-2897.
%
\bibitem{shilnikov2005}
A. Shilnikov, G. Cymbalyuk, Phys. Rev. Lett \textbf{94}, (2005) 048101.
%
\bibitem{han2016}
X. Han, F. Xia, P. Ji, Q. Bi, J. Kurths, Commun. Nonlinear Sci. Numer. Simul \textbf{36}, (2016) 517-527.
%
\bibitem{yu2016}
Y. Yu, Z. Zhang, Q. Bi, Y. Gao,  Appl. Math. Model \textbf{40},  (2016) 1816-1824.
%
\bibitem{wang2020}
Z. Wang, Z. Zhang, Q. Bi, Nonlin. Dyn \textbf{100}, (2020) 2899-2915. 
%
\bibitem{han2017c}
X. Han, F. Xia, C. Zhang, Y. Yu, Nonlin. Dyn \textbf{88},(2017)  2693-2703.
%
\bibitem{han2018}
X. Han, Q. Bi, J. Kurths, J, Phys. Rev. E \textbf{98}, (2018) 010201(R).
%
\bibitem{wei2021}
M. Wei, W. Jiang, X. Ma, X. Han, Q. Bi, Nonlin. Dyn \textbf{104}, (2021) 4493-4503.
%
\bibitem{ma2021}
X. Ma, W. Jiang, Y.Yu,  Commun. Nonlinear Sci. Numer. Simul \textbf{103}, (2021) 105959.
%
\bibitem{vo2016}
 T. Vo, M. A. Kramer ,T. J. Kaper, Phys. Rev. Lett \textbf{117}, (2018) 268101.
 \bibitem{ryashko2017}
 L. Ryashko, E. Slepukhina, Phys. Rev. E \textbf{96}, (2017) 032212.
 \bibitem{han2017d}
X. Han , Y. Yu ,  C. Zhang, F. Xia , Q. Bi, Int. J. Nonlin. Mech \textbf{89}, (2017) 69-74.
\bibitem{chen2017}
Z. Chen, F. Chen, Nonlin. Dyn \textbf{100}, (2017) 659-667.
\bibitem{song2020}
J. Song, M. K. Wei, W. A. Jiang, X.F. Zhang, X. J. Han, Q.S. Bi, Acta Phys. Sin \textbf{69}, (2020) 070501.
\bibitem{chen2020}
Z.Y. Chen , F.Q. Chen, Chaos Soliton Fract \textbf{137}, (2020) 109814.
\bibitem{buchele2006}
B. B\"uchele , H. Kreibich , A. Kron , A. Thieken , J. Ihringer , P. Oberle, B. Merz, F. Nestmann, Nat. Hazards Earth Syst. Sci \textbf{6}, (2006) 485-503.
\bibitem{sachs2012}
M. Sachs, M. Yoder, D. Turcotte, J. Rundle, B. Malamud,  Eur Phys J Spec Top \textbf{205}, (2012) 167-182.
\bibitem{bird2011}
D. K. Bird, C. C. Goff, A. Gero, Aust. Geogr \textbf{42}, (2011) 225-239.
\bibitem{hoerling2013}
M. Hoerling, A. Kumar, R. Dole, J. W. Nielsen-Gammon, J. Eischeid, J. Perlwitz, X-W Quan , T. Zhang, P. Pegion, M. Chen, Anatomy of an extreme event. J. Clim \textbf{26}, (2013) 2811-2832.
\bibitem{scheffer2003}
M. Scheffer, S. R. Carpenter, Trends Ecol. Evol \textbf{18}, (2003) 648-656.
\bibitem{mcmichael2015}
A.J. McMichael, Virulence \textbf{6}, (2015) 543-547.
\bibitem{machado2020}
J. A. T. Machado , A.M. Lopes, Nonlin. Dyn \textbf{100}, (2020) 2953-2972.
\bibitem{lehnertz2006}
K. Lehnertz, Epilepsy: extreme events in the human brain, in: Extreme events in nature and society (Springer, Berlin 2006) 123-143.
\bibitem{frolov2019}
N. S. Frolov, V.V Grubov, V. Maksimenko , V.V L\"uttjohann Makarov, A.N. Pavlov, E. Sitnikova, A.N. Pisarchik, J. Kurths, A. E. Hramov, Sci. Rep \textbf{9}, (2019) 1-8.
\bibitem{buzulukova2017}
N. Buzulukova, Extreme events in geospace: Origins, predictability, and consequences.  (Elsevier 2017) 123-143.
\bibitem{krause2015}
S.M. Krause, S. B\"orries, S. Bornholdt, Phys. Rev. E \textbf{92}, (2015) 012815.
\bibitem{helbing2001}
D. Helbing, Rev. Mod. Phys \textbf{73}, (2001) 1067.
\bibitem{zhao2005}
L. Zhao, Y-C. Lai, K. Park, N. Ye,  Phys. Rev. E \textbf{71}, (2005) 026125.  
\bibitem{chen2015}
Y-Z. Chen, Z-G. Huang, H. F. Zhang, D. Eisenberg, T. P. Seager, Y-C. Lai, Sci. Rep \textbf{5}, (2015) 17277.
\bibitem{dobson2007}
I. Dobson, B. A. Carreras, V.E. Lynch, D.E. Newman, Chaos \textbf{17}, (2007) 026103.
\bibitem{almarwae2017}
M. Almarwae, Sci. World J \textbf{66}, (2017) 97.
\bibitem{suresh2021}
R. Suresh, R. Gopal, S. Nataraja Pillai, V. K. Chandrasekar, Evidence of extreme events in wind-induced normal stress of the columns of low- and medium-rise building structures (unpublished) 
\bibitem{chowdhury2021}
S.N. Chowdhury, A. Ray, S.K. Dana, D. Ghosh. Extreme events in dynamical systems and random walkers: A review. arXiv Preprint arXiv:2109.11219 (2021).
\bibitem{zamora2013}
J. Zamota-Munt, B. Garbin, S. Barland, M. Giudici, J.R.R. Leite, C. Masoller, J.R. Tredicce,  Phys. Rev. A \textbf{87}, (2013) 035802.
%
\bibitem{ray2019}
A. Ray, S. Rakshit, D. Ghosh, S.K. Dana, Chaos \textbf{29}, (2019) 04313.
\bibitem{bonatto2017}
C. Bonatto, A. Endler, Phys., Rev. E \textbf{96}, (2017) 012216.
\bibitem{kaviya2020}
B. Kaviya, R. Suresh, V.K. Chandrasekar, B. Balachandran, Int. J. Nonlin. Mech \textbf{127}, (2020) 103596.
%
\bibitem{suresh2020}
R. Suresh, V.K. Chandrasekar, Chaos \textbf{30}, (2020) 083141.
%
\bibitem{suresh2018}
R. Suresh, V.K. Chandrasekar, Phys. Rev. E \textbf{98}, (2018) 052211.
%
\bibitem{kingston2017}
S. Leo Kingston, K. Tamilmaran, P. Pal, U. Feudel, S. K. Dana, Phys. Rev. E \textbf{96}, (2017) 052204.
%
\bibitem{kumarasamy2018}
S. Kumarasamy, A. N. Pisarchik, Phys. Rev. E \textbf{98}, (2018) 032203.
%
\bibitem{mishra2018}
A. Mishra, S. Saha, M. Vigneshwaran, P. Pal, T. Kapitaniak, S. K. Dana, Phys. Rev. E \textbf{97}, (2018) 062311.
%
\bibitem{chowdhury2019}
S.N. Chowdhury, S. Majhi, M. Ozer, D. Ghosh, M. Perk, New. J. Phys \textbf{21}, (2019) 073048.
%
\bibitem{pisarchik2011}
A. N. Pisarchik, R. Jaimes-Re\'ategui, R. Sevilla-Escoboza, G. Huerta-Cuellar, M. Taki, Phys. Rev. Lett \textbf{107}, (2011) 274101.
\bibitem{pisarchik2012}
A.N. Pisarchik, R. Jaimes-Re\'ategui, R. Sevilla-Escoboza, G. Huerta-Cuellar, Phys. Rev. E  \textbf{86}, (2012) 056219.
%
\bibitem{kanai2012}
Y. Kanai, H. Yabuno, Nonlinear Dyn. \textbf{70}, (2012) 1007.
%
\bibitem{erlicher2013}
S. Erlicher, A. Trovato, P. Argoul, Mech. Syst. Signal Process. \textbf{41}, (2013) 485.
%
\bibitem{warminski2010}
J. Warmi\'nski, Nonlinear Dyn. \textbf{61}, (2010) 677.
%
\bibitem{szabelski1995}
K. Szabelski, J. Warmi\'nski, Int. J. Nonlinear Mech. \textbf{30}, (1995) 179.
%
\bibitem{hasegawa2011}
H. Hasegawa, Phys. Rev. E \textbf{84}, (2011) 061112.
%
\bibitem{miwadinou2018}
C. H. Miwadinou, A. V. Monwanou, A. A. Koukpemedji, Y. J. F. Kpomahou, J. B. Chabi Orou, Int. J. Bifur. Chaos \textbf{28}, (2018) 1830005.
%


\bibitem{margallo1992}
J.G. Margallo,  J.D. Bejarano,  J. Sound Vib \textbf{156}, (1992) 283-301.
\bibitem{maccari2001}
A. Maccari, Nonlin. Dyn \textbf{25}, (2001) 293-316.
\bibitem{kpomahou2021}
Y.J. F. Kpomahou, L.A. Hinvi, J.A. Ad\'echinan, C.H. Miwadinou, Complexity, (2021) 6631094.
\bibitem{eichler2011}
A. Eichler, J.  Moser, J. Chaste, M. Zrdojek, I. Wilson-Rae, A. Bachtold, Nat. Nanotechnol \textbf{6}, (2011) 339-342.
\bibitem{zaitsev2012}
S. Zaitsev, O. Shtempluck, E. Buks, O. Gottlieb, Nonlin. Dyn \textbf{67}, (2012) 859-883.
\bibitem{ran2009}
Z. Ran, Appl. Fluid. Mech \textbf{5}, (2009) 41-67.
\bibitem{ekinci2004}
K.L. Ekinci, Y.T. Yang, M.L. Roukes,  J. Appl. Phys \textbf{95}, (2004) 2682-2689.
\bibitem{papariello2016}
L. Papariello, O. Zilberberg, A. Eichler, A. Chitra,  Phys. Rev. E \textbf{94}, (2016) 022201. 
\bibitem{akerman2010}
N. Akerman, S. Kotler, Y. Glickman, Y. Dallal, A. Keselman, R. Ozeri, Phys. Rev. A \textbf{82}, (2010) 061402(R).
\bibitem{han2018a}
X. Han, Y. Zhang, Q. Bi, J. Kurths, Chaos \textbf{28}, (2018) 043111. 
%
%
  
 
%
     
          
 


\end{thebibliography}

\end{document}